# First passage time and change of entropy

V. V. Ryazanov
Institute for Nuclear Research, pr. Nauki, 47 Kiev, Ukraine,
e-mail: vryazan19@gmail.com

**Abstract.** The first-passage time is proposed as an independent thermodynamic parameter of the statistical distribution that generalizes the Gibbs distribution. The theory does not include the determination of the first passage statistics itself. A random process is set that describes a physical phenomenon. The first passage statistics is determined from this random process. The thermodynamic parameter conjugated to the first-passage time is the same as the Laplace transform parameter of the first-passage time distribution in the partition function. The corresponding partition function is divided into multipliers, one of which is associated with the equilibrium parameters, and the second one - with the parameters of the first-passage time distribution. The thermodynamic parameter conjugated to the first-passage time can be expressed in terms of the deviation of the entropy from the equilibrium value. Thus, all moments of the distribution of the first passage time are expressed in terms of the deviation of the entropy from its equilibrium value and the external forces acting on the system. By changing the thermodynamic forces, it is possible to change the first passage time.

## 1. Introduction

The first-passage time (*FPT*) is the time during which a stochastic process reaches a certain threshold for the first time. *FPT* distributions have found numerous applications in physics, biology and finance science [1–3]. Examples of the use of the *FPT* parameter include frequency-tuning systems [4], problems of crossing a potential barrier, decay of an unstable state [5], conformational changes of proteins, ecological systems, epidemics, diffusion-controlled chemical reaction, autocatalytic reactions, number of individuals in a population, dissociation of a diatomic molecule, potential of a neuron, transition through a potential barrier [6]. In [7] the *FPT*, the moment of the first reach of a certain threshold by a random process, was used to study the effects of noise on electronic relays, and for other problems of radio engineering. In [8, 9] the possibilities of applying distributions of *FPT* (referred to as "lifetime" in [7]) to the description of superstatistics and such phenomena as the behavior of systems with multiplicative noise, Van der Pol-Duffing systems, turbulence, Van der Pol generators are noted. The *FPT* distribution has been studied for a variety of diffusive processes, ranging from ordinary diffusion (Brownian motion) to continuous-time random walks [10–12], and in many other applications [1-3]. First-passage-time distributions were studied in the context of thermodynamics as well [13, 14–19].

A large variety of problems ranging from noise in vacuum tubes, chemical reactions and nucleation [19] to stochastic resonance [20], behavior of neurons [21], and risk management in finance [1] can be reduced to *FPT* problems.

A Gaussian approximation was used in [5] in the problem of transition through a potential barrier. For other processes, other expressions are possible for the probability density of the first passage of a level. In [22], for the probability density of the first passage of the level in the case of Brownian motion, an inverse Gaussian distribution density was obtained depending on the boundary conditions and on the time dependence of the drift and diffusion coefficients. In [23], a generalization of the Eyring – Kramers transition rate formula for irreversible diffusion processes was obtained. In [24] the Weibull distribution was derived for the probability density of the first passage of a level. In [25], the escape probability was obtained for the Feller process. In [26], the limiting exponential distribution for this value was derived. In [27] the authors provided an exact unified framework for studying the comprehensive statistics of first passage time under detailed balance conditions. In [28] an exact determination of the first passage time distribution from the corresponding relaxation spectrum was performed. In [29] the behavior of smooth *FPT* densities with finite moments was explained.

In [30-33] *FPT* is determined for the diffusion processes as a time it takes for the fastest "diffusive searcher" to find a target from a large group of searchers. In [34] a framework is set for analyzing extreme value statistics of ergodic reversible Markov processes in confining potentials on the hand of the underlying relaxation eigen spectra. New trends in first-passage methods and overview of *FPT* problems can be found in [35].

It is of interest to study the possibilities of controls over *FPT*. In [36, 37, 38], the possibilities of the optimal control problem are considered. In [13] the steady state entropy production rate was estimated using mean first-passage times of suitable physical variables. In the general case, the impact on a physical system is described by the flows into the system from the environment and by the corresponding thermodynamic forces or by the deviation of the entropy of the system from its equilibrium value. This deviation consists of the internal production of entropy, expressed through flows and forces, and the flow of entropy into the system from the outside. The main goal of this work is to establish a relationship between the *FPT* distribution parameters and the deviation of the entropy of the system from its equilibrium value. Revealing such relations makes it possible to affect *FPT* by various external controls. The possibility of such controlling depends on the type of *FPT* distribution. For large times, the limiting distribution of *FPT* is exponential. Then any impacts (if they act over a sufficiently long period of time) can only decrease the mean *FPT*. For other *FPT* distributions, which are possible for shorter periods of time of impact on the system (for example, [39]), there are means to increase the mean *FPT*. The finite values of *FPT* are considered, and the distributions of these values should describe finite, and not only infinitely large values of the exposure time.

In this paper, *FPT* is considered as an independent thermodynamic variable [8, 9, 40, 41, 42]. Similar thermodynamic parameters of flows are considered in the extended irreversible thermodynamics (*EIT*) [43]. The distribution containing *FPT* includes the conjugate thermodynamic parameter $\gamma$. It is expressed through the deviation of entropy from the equilibrium value $\Delta s$. The same parameter $\gamma$ appears in the *FPT* Laplace transform. Thus, the mean value and all further moments of *FPT* are expressed in terms of $\Delta s$. Distributions generalizing Gibbs distributions by introducing an additional parameter, generalized or canonical Gibbs reservoirs, which are thermodynamic reservoirs in a field of conservative external forces, are discussed in [44].

There are known relationships between the average time of crossing the barrier and the entropy of the transition in the reaction-rate theory and escape from metastable states (for example, [5, 28]). In [45] an expression was obtained for the transition time scale, which coincides with the expression from [40-41] for the average lifetime (in terms of [7]) and the exponential distribution of the lifetime after changing the sign of the entropy change. Here we consider a more general case.

The relationship between time and entropy has been known for a long time (for example, [46]). In [16, 17, 47] *FPT*s are considered for a certain level of entropy production or for a given flow value. In [48] it is shown that the rate of dissipation $\dot{s}$ in a nonequilibrium stationary process *X* bounds the moments of *FPT* of dissipative currents *J*. In [49] for the birth and death process, a result obtained for the relationship between the average *FPT* and the change in entropy, which corresponds to the result obtained for the continuous exponential distribution in [40-41]. In this paper we obtain the relation between the *FTP*, the value directly related to the passage of time, and thermodynamic quantities.

In this paper, an independent thermodynamic variable *FPT* is included in the statistical distribution [8, 9, 42]. In this case, *FPT* is considered not as a temporary variable, but as a quantity that depends on the phase coordinates of the system. Thermodynamics and first passage times both are important fields of study. However, there is a certain gap between them. First passage is a concept from the stochastic theory. The approach proposed in this paper allows us to bring these two important areas closer together, and such a rapprochement would mutually enrich them. Relationships are obtained between the movement, rates of the processes in the system and thermodynamic forces and flows in it. The assumption that *FPT* can be an independent thermodynamic parameter makes it possible to relate *FPT* to the change in entropy of the system and to combine many stochastic relations for *FPT* with the physical results of non-equilibrium and stochastic thermodynamics.

In the second Section, the introduction of *FPT* as a thermodynamic parameter is substantiated. The generalizing Gibbs distribution containing the energy and *FPT* and their conjugate parameters is introduced. The third Section deals with generalized thermodynamics of systems with an extra parameter of *FPT*. The entropy and its derivatives are determined both for the distribution depending on the energy $u$ and the *FPT* $T_\gamma$, when the entropy is equal to $s$ (12), and for the distribution depending only on $y=u$. The fourth Section discusses the *FPT* of a certain flow level and in a drift-diffusion process. Explicit dependences of the parameter $\gamma$, which is the thermodynamically conjugated *FPT*, through changes of entropy are obtained. The proposed theory does not allow to determine first passage statistics itself. General relations are obtained that are valid for arbitrary *FPT* distributions. The choice and use of specific *FPT* distributions depends on the physical situation under study. The fifth section contains some concluding remarks of the paper. Appendix A discusses the Feller process, for which some controls can be used to increase the mean *FPT*.

## 2. Distributions for energy and *FPT*

The initial values of the phase coordinates are unknown. Therefore, statistical ensembles are introduced. Thus, a probabilistic interpretation of dynamic processes is assumed. The parameters of the energy $u$ and *FPT* $T_\gamma$ (1) are considered as random functions in the phase space. Random variable $T_\gamma$ (1) is considered as random variable in the phase space, presenting a numerical function that define a measurable mapping of the phase space $z$ to the real number space. In the reduced description, the kinetic equations for the distribution functions are derived from the Liouville equation [50]. The values of internal energy $u$ and *FPT* $T_\gamma$ are described by direct and inverse conjugate kinetic equations [4].

The inclusion of *FPT* in the category of thermodynamic variables allows to introduce a statistical distribution containing *FPT*, a modified entropy, and relate the change in entropy to *FPT*. This opens wide possibilities for studying *FPT* from the standpoint of statistical thermodynamics, searching for and studying numerous connections between stochastic and thermodynamic quantities and regularities. The processes of reaching a certain level are accompanied by a change in entropy, which must be taken into account. Let us point out why *FPT* can be considered as a thermodynamic parameter. First of all, *FPT* depends on phase coordinates and corresponds to the definition of thermodynamic parameters given in [51, 52]. The kinetic equations for the distributions of internal energy and *FPT* are conjugated (these are "forward" and "backward", e. g. Fokker-Planck equations). Therefore, *FPT* is as essential to statistical thermodynamics as internal energy, which is the basic value in statistical physics. Examples of the use of thermodynamic parameters similar to *FPT* are flows in the *EIT* and the random moment of system birth, the random time of the system's past lifetime in the Non-equilibrium Statistical Operator (*NSO*) method [53, 54]. Fluxes have a lot in common with *FPT*. In some cases, the flow is inversely proportional to *FPT*.

*FPT* was studied mainly from the point of view of s stochastic processes, an important and necessary area of mathematics, to which *FPT*'s belong [1-7, 13-18, 22-39, 45, 47-49, 56, 57]. This should be considered as an indispensable stage for further research, for example, where thermodynamic aspects of *FPT* are considered. In this paper, the relations of *FPT* with statistical thermodynamics are established. The results of stochastic treatment of *FPT*, for example, the Laplace transform of the distribution density of *FPT* are exemplified.

The main goal of this paper is to explicitly relate a random *FPT* process (stochastic process reaching a preset level) to the changes in the entropy of the system that occur during this process. In any real physical process modeled by some random mathematical process, the entropy changes that accompany the process are inevitable. They should be taken into account when correctly describing the *FPT* process. During the *FPT* process, the entropy of the system changes. These changes should be taken into account in the moments of *FPT* distribution. *FPT*'s are a special case of stopping times [56, 57]. In this paper, a joint distribution for internal energy and *FPT* is obtained. Instead of internal energy, another quantity can be considered, for example, neutron flux density in a nuclear reactor. The introduced distribution containing *FPT* as a thermodynamic parameter is compared with the non-equilibrium distribution in the *NSO* approach by Zubarev [53, 54]. The case is considered when the distribution of the random *FPT* does not depend on the random value of the internal energy $u$, but rather depends on the average values of the internal energy. In the model used in this work and the approximations used in this case, the Laplace transform of the *FPT* distribution appears, the argument of which is associated with the change in the entropy of the system using thermodynamic relations. Therefore, the value, for example, of the average *FPT* value, in which the value of the argument of the Laplace transforms of the *FPT* distribution assumed zero, is just an auxiliary value that does not correspond to real processes in which the changes in entropy and the values of this argument are not zero. It is necessary to set the value of the argument of the Laplace transform, which corresponds to the change in entropy in the real *FPT* process of reaching the specified level.

The change in entropy, obtained using the introduced distributions, is expressed in terms of thermodynamic flows and conjugate thermodynamic forces. There is also an inverse relationship: the effect of entropy changes caused by the introduction or change of, for example, thermodynamic forces on the *FPT*. We can study the problem of controlling *FPT*. The system can be affected by various kinds of thermodynamic forces that cause thermodynamic flows and entropy changes, all affecting the *FPT*. Using the obtained relationship between the average *FPT* and the change in entropy, the intervals of entropy change are determined, at which the average *FPT* of reaching the level decreases or increases. Processes in the system slow down or speed up.

An important problem is the possibility to control and change the properties of the *FTP* processes. The proposed approach suggests a solution to this problem using the methods of the theory of random processes, non-equilibrium statistical physics and thermodynamics. The time of the first passage (hitting a preset level by a stochastic process) is set as a thermodynamic parameter, and a joint distribution is constructed for the energy of the system and the time of the first passage, then an expression is written for the non-equilibrium entropy of the system. By changing the thermodynamic forces, it is possible to change the entropy of the system and the time of the first passage. The system is open, and the parameters of the system under the impact of the environment are random variables that are characterized stochastically. In the present work, as in [8, 9, 41, 42, 43], an additional thermodynamic parameter is introduced to describe a non-equilibrium state.

*FPT* is the time it takes for a stochastic process $X(t)$ to reach a certain threshold $a$ (1) for the first time. For example, it is the time for a stochastic process that describes a macroscopic parameter $X(t)$ to reach the zero level. The stochastic process $X(t)$ is the order parameter of the system, for example, its energy, number of particles (in some papers, for example [47], entropy production is considered). The process $X(t)$ can describe many other physical quantities. *FPT* is, by definition,

$$T_{\gamma x} = \inf\{t : X(t) = a\}, \quad X(0) = x > 0. \tag{1}$$

The subscript $\gamma$, emphasizing the dependence on the conjugate thermodynamic parameter, is used not to confuse this variable with the temperature *T*. There are other definitions of *FPT*. For example, $T_{\gamma x} = \inf\{t > 0; X(t) \in A\}$, where $A$ is a Borel set on a number line, $T_{\gamma x}$ is the moment of the first achievement of the set *A* [56]. The moments of the first passage refer to the stopping times and Markov moments [56, 57]. Each Markov moment is associated with a number of sets describing a set of events observed over a random time $T_{\gamma x}$ [57]. Physically, this corresponds to a dependency of the system on the system history. The events that occur during this time include the change in the entropy of the system. In this way, the system history is taken into account. *FPT* (1) is a multiplicative functional from a random process *X(t)* [56]. In a joint distribution for internal energy and *FPT*, as in the *NSO* method, the dependence on the system past is important.

In [8, 9, 42] a distribution is introduced that contains *FPT* (referred to as "lifetime") as an additional thermodynamic parameter. The choice of *FPT* as a thermodynamic parameter is possible from the mere definition of a thermodynamic parameter in statistical physics. For example, in [51] and [52] it is stated: "Any function *B(z)* of dynamic variables $z=(q_1,...,q_N, p_1,...,p_N)$ of macroscopic nature is a random inner thermodynamic parameter". The fact that *FPT* (1) depends on dynamic variables *z* can be seen from the equations for the distribution of the *FPT* (or lifetime [7]) in the Markov model [4, 56]. *FTP* is considered in the statistical ensemble not as a time variable, but as a parameter that depends on the phase coordinates *z*.

The distributions for *FPT* (1) are described by the Pontryagin equations [4], [56], which are conjugate to the kinetic equations (for example, the Fokker-Planck equations) for the energy distribution of the system, the main variable of statistical thermodynamics and equilibrium statistical physics. Since the Fokker-Planck equation contains functions that depend on phase variables, the same dependences hold for the conjugate equation.

A plausible illustration is the method of the non-equilibrium statistical operator (*NSO*). *FPT* is contained in the *NSO* method implicitly. In *NSO* [53]-[54], [59, 60, 61, 62] an additional macroscopic parameter in the description of the non-equilibrium system is the time $t-t_0$ elapsed from the birth of the system, the time of the past life, the time until the first crossing of the zero level in the inverse time, or the age of the system [55]. In [55] it was shown that averaging over the initial time [53-54] corresponds to averaging over the distribution of the system lifetime $t-t_0$. We consider the classical case. However, as noted in [54], the term operator is used for both the classical and quantum cases. The *NSO* method is associated with the distribution introduced in this paper.

In [53, 54] the limit transition $t_0 \rightarrow -\infty$, $t_0-t \rightarrow -\infty$, is made, after taking the thermodynamic limit. Before this transition is made, the functions in *NSO* depend on the value $s=t-t_0$, which in [55] is interpreted as the random lifetime of the system, *FPT*. Zubarev's approach follows from the choice of the weight function *w(t, t')* introducing in the integration in time [60, 61, 62] in the form of Abel kernel $\varepsilon exp\{\varepsilon(t'-t)\}$ and $t_0$ taken in the remote past; $t_0$ is the initial time of preparation. In [55] it is shown that the logarithm of *NSO* is equal the average from the logarithm of relevant (or quasi-equilibrium) distribution on the system lifetime distribution.

This is an introduction of an *ad hoc* time-smoothing procedure, with properties, which allow to construct a consistent non-equilibrium statistical operator. In particular, its fading-memory property leads to the irreversible evolution of the macrostate of the system, from the initial condition of preparation (the information available to build the formalism) and the description of the system on the basis of such information and its evolution to a final state of equilibrium in the future.

The *NSO* dependences on $s=t-t_0$ differ before and after the limit transition $t_0 \rightarrow -\infty$. So, *NSO* depends on the random variable $s=t-t_0$, since $t_0$ is a random variable. After the limit transition, the dependence on *s* becomes integrated over *s*, averaging occurs.

By definition (1) *FPT* is the first moment when a random process *X(t)* reaches the absorbing set "*a*". *FPT* $T_\gamma$ plays the role of a subordinate process [63]. In *NSO* the absorbing set "*a = 0*" is achieved in the reverse time. If in *NSO* method we consider the process in reverse time after the replacement $t \rightarrow -t$, then the birth of the system at a random moment $t_0$ corresponds to its death at this random moment. The process *X(t)* from expression (1) in direct time describes the random process of the system's existence, the time of the system's past life, from the random moment $t_0$ to the current moment *t*. Achievement of a zero value in the reverse time at a random moment $t_0$ corresponds to the *FPT* process reaching the limit $a=0$ specified in expression (1) by a random process for the number of particles in the system or the energy of the system; the role of a random process *X(t)* is played by the dynamical variables $\hat{P}_m$. Parameters $\hat{P}_m$ start from the current moment *t*, and at a random moment $t_0$ in the reverse time reach the absorbing set "$a=0$".

For finite-size systems with finite times of reaching the zero level by the number of particles and energy, lifetime, a special case of *FTP*, in the Liouville equation there is a source that characterizes the interaction of finite-size system with the environment. For finite lifetimes of the system, the source in the Liouville equation is not zero. For the exponential distribution of time $t-t_0$, $w(t,t_0)=(1/<M>)exp[-(t-t_0)/<M>]$ [53, 54], the value $<M>=\varepsilon^{-1}=<t-t_0>$ is interpreted in [55] as the average lifetime of the finite non-equilibrium system, and this source stands in the right-hand side of the Liouville equation. It is also possible to set an expression for the distribution of time $t-t_0$ other than exponential. Examples of such a choice are given in [42, 64]. The lifetime distribution may differ significantly from the limiting exponential distribution [42, 64].

Section 4 considers an example of *FPT* for a diffusion process with an absorbing boundary. In this ensemble, particles reaching the boundary are absorbed there and removed from the system. Another example is overcoming some energy barrier, for example, in chemical reactions. The system is open, hence it can also contain some incoming particle flows. The balance between the input and output determines the lifetime of the system, as the random moment of degeneration of the system when there are no particles left. The ensemble is characterized by varying initial values of the phase coordinates $(z_0=q_1(t=0),...,q_{N(t=0)}(t=0), p_1(t=0),..., p_{N(t=0)}(t=0))$. Energy and *FPT* are functions of phase coordinates. Therefore, in such an ensemble, they take random values with different probabilities, and cover the entire space of possible values. Below, in relations (6) - (9), as in [8, 9, 40-42], a generalization of the canonical ensemble is introduced. In the example considered, the number of particles is variable, and a generalization of the grand canonical ensemble must be considered. It is also possible to use the property of the thermodynamic equivalence of ensembles.

In non-equilibrium systems, in contrast to equilibrium ones, time scales are essential. The time when an event happens (for example the escape of a particle from a metastable state) is modeled with a stopping time $T_\gamma$ [47, 56, 57]. These include *FPT*.

Reduced description of non-equilibrium systems [54, 59] is a description based on the restriction information about the system. In this case, the difference between the spatial and temporal scales of the processes occurring in non-equilibrium systems is used.

There are several different ways to construct the distribution function of thermodynamic parameters [53, 65, 66]. To determine the distribution density for energy and *FPT* $\rho(z; u, T_\gamma)$, we will proceed from the "objective" school regarding the probability of an event as an objective property of that event, always capable in principle of empirical measurements by observation of frequency ratios in a random experiment [67], based on mathematical statistics. Maximum entropy method [67], [53] is also suitable for the derivation of the distributions (7), (8).

Statistical distribution functions are considered in many sources (for example, [53, 65, 66]). We turn from the discussion of a single system to that of a set of similar systems. To this end, let us assume that initially we have *N* identical systems. If we subject them to some realistic preparation procedure, the outcome will be a set of *N* slightly different systems, which we call an ensemble. Hence it makes sense to describe an ensemble in *2Nd*-dimensional phase space of sufficiently many systems by a continuous function $\rho(p, q, t)$ defined by

$\forall B \subset \Gamma : \int_B \rho(p,q,t)d^{dN}p\, d^{dN}q \propto$ number of ensemble points in the set *B* and the normalization $\int_\Gamma \rho(p,q,t)d^{dN}p\, d^{dN}q = 1$; $\Gamma$ is phase space of the position and momentum coordinates of the particles in *d* dimensions. It is easy to see that the $\rho$ thus defined is just a probability density in phase space $z=q_1,...,q_N, p_1,...,p_N$. Function $\rho(p,q,t)=\rho(z,t)$ therefore is called a macrostate of classical mechanics. Let $\rho$ be a macrostate and $A = A(p,q)$ an observable.

The probability density $p_\rho$ of an observable $A$ for a macrostate $\rho$ given by the expression [65], [66]:

$$p_\rho(A) \equiv \rho(A) = \int_\Gamma \delta(A - A(p,q))\rho(p,q,t)d^{dN}p\,d^{dN}q = \langle \delta(A - A(p,q)) \rangle_\rho. \qquad (2)$$

where $\rho(z;t)$ is the density of distribution of coordinates $q_i$ and momenta $p_i$ of particles of the system; $i=1,2,...,N$. The function $\rho(p,q,t)$ is also determined by the residence time of the system in a certain region of the phase space [51].

In [53, 54, 59] generalized Gibbs ensembles are constructed, which are closely related to the thermodynamic description of non-equilibrium systems, when the observable macroscopic quantities depend on time. These generalized Gibbs ensembles are also called the relevant ensembles or quasi-equilibrium ensembles. The relevant statistical distributions serve as auxiliary distributions to select spatial solutions of the Liouville equation that describe irreversible macroscopic processes [60-62]. The non-equilibrium macroscopic state is specified by the set of observables which are the average values $<\hat{P}_m>^t$ of some relevant dynamical variables $\hat{P}_m$. These variables give the reduced description of the system on the selected time scale. The existence of different time scales is due to the "hierarchy" of basic relaxation times in macroscopic systems. Using the Lagrange multipliers method [53], [54] (or other approaches, such as mathematical statistics, section 2), it follows that

$$\rho_{rel}(t,0) = \exp[-\Phi(t) - \sum_m F_m(t)\hat{P}_m], \qquad (3)$$

which will be called the "relevant distribution". Here a set of variables $\{F_m(t)\}$ play the role of variables thermodynamically conjugated to the macrovariables $\hat{P}_m$. The Massieu-Planck function $\Phi(t)$ is determined from the normalization condition for the relevant distribution (3).

We proceed to an reduced description for the distribution density $\rho(z;t)$. In this case, as in [53], [59], we seek solutions of the Liouville equation that depend on time only through the mean values of a certain set of observables $<\hat{P}_m>^t$. The distribution density $\rho(z;t)$ from expression (2) is replaced by a distribution of the form (3) $\rho_{rel}(z;u,T_\gamma)$.

*FTP* (1) in terms of the stochastic theory [57, 63] is a functional of the main random process $X(t)$ (for example, energy), and a subordinate random process. *FTP* is a macroscopic and slowly changing quantity. The distribution for the *FPT* $T_{\gamma x}$ (1) depends on the macroscopic values of $X(t)$. Assume that the process $X(t)=u$ is the energy (or another system order parameter) of the system. The relationship between the distribution density $p(u,T_\gamma)=p_{uT_\gamma}(x,y)$ and the microscopic (coarse-grained) density $\rho_{rel}(z;u,T_\gamma)$ is written as (following the standard procedure, for example, see [65], [66]):

$$p(u,T_\gamma) = \int \delta(u - u(z))\delta(T_\gamma - T_\gamma(z))\rho_{rel}(z;u,T_\gamma)dz. \qquad (4)$$

In the case of deterministic chaos, the function $\delta(A-A(z))$ ($A=u, T_\gamma$) from (4) is replaced by $\delta(A-f(A(z)))$, where $f$ is a mapping (a chaotic map is a map (=evolution function) that exhibits some sort of chaotic behavior) [68]. However, as it is written in [47]: "We say that a random time $T \in [0,\infty) \cup \{+\infty\}$ is a stopping time if it is a deterministic function defined on the set of trajectories $X_0^{+\infty} = \{X(t)\}_{t \in R^+}$ that obeys causality; in other words, the value of the stopping time $T$ is independent of the outcomes of the process $X$ after the stopping time. If the event does not occur, then $T = +\infty$". In this case and for the case of stochastic dynamics in (4) the expression $\delta(T_\gamma-T_\gamma(z))$ is valid. We consider the observable quantity, the thermodynamic parameter of *FPT* $T_\gamma$, as random variable in the phase space $z$ (due to the uncertainty of the initial values of the phase coordinates) and deterministic function in the time dynamics. The particles obey the equations of motion of mechanics, $T_\gamma$ is described by the Hamilton equations.

There are different possibilities for determining the function $\rho_{rel}(z;u,T_\gamma)$ from expression (4). For example, in [53] physical quantities characterizing the system $\xi_1,...,\xi_k$ are specified, but not necessarily the integrals of motion. Average values $\langle \xi_k \rangle$ can characterize some state of incomplete statistical equilibrium. The free energy of a non-equilibrium state, characterized by a given mean value $\langle \xi_k \rangle$, is defined in [51] as the free energy of an equilibrium state in auxiliary fields that make the system in equilibrium at given values $\langle \xi_k \rangle$. In this case, the distribution density is

$$\rho_{rel} = Q_p^{-1} \exp[-\beta(H - \mu N - \sum_k a_k \xi_k)],$$

where $\beta$ is the inverse temperature, $\mu$ is the chemical potential, $H$ is the energy, $N$ is the number of particles. The partition function $Q_P$ is a function of $\beta, \mu, a_k$ or $\beta, \mu, \langle \xi_k \rangle$ and determines the thermodynamic functions in a state of incomplete statistical equilibrium with a given $\langle \xi_k \rangle$ depending on $\beta, \mu, \langle \xi_k \rangle$. We put $\xi_1 = T_\gamma$; the auxiliary field will be described by the parameter $a_1 = \gamma$ from expression (8). In [53], this approach is also used in the local equilibrium distribution, a particular case of the relevant distribution [54]. This description is valid for not very large deviations from equilibrium. Similar relations were written in [69] for the generalized Gibbs ensemble, which corresponds to a given type of thermodynamic contact of a macroscopic system with the environment.

Let us consider another approach to deriving the function $\rho_{rel}(z;u,T_\gamma)$ from Eq. (4). It is based on the general results of mathematical statistics and is free from limitations.

It was shown in [70] that for a sequence of independent trials with $n$ outcomes, each with probability $p_i$, $\ln P\{(v_1,...,v_n) \in W\} \approx -ND(P|R)$, where $R$ is the maximum point of $D(P\|Q)$ at $Q \in W$, $v_1,...,v_n$ are the frequency of outcomes, the region $W$ satisfies the requirements of regularity. Information deviation (or relative entropy) $D(P\|Q)$ of the probability distribution $Q$ from the dominated distribution $P$, $Q \gg P$ is the quantity $D(P\|Q) = \begin{cases} \int \log(dP/dQ)dP, & \text{if } P \ll Q \\ +\infty, & \text{else} \end{cases}$.

The notation $\lambda \gg \mu$ denotes that the measure $\lambda$ dominates $\mu$, $Z_\mu \subseteq Z_\lambda$, where $Z_\mu$ is the ideal of the algebra $S$ of measure $\mu\{\circ\}$ on a measurable space $(\Omega, S)$, all sets of zero measure, zero sets, $\mu\{\{\omega\} \subseteq Z_\mu\} = 0$ [71]. By the Radon-Nikodym theorem, if $Z_\mu \subseteq Z_\lambda$, then on $\Omega$ there is a derivative $\frac{d\nu}{d\mu}(\omega)$ of the measure $\nu$ with respect to measure $\mu$, a measurable finite numerical function $h(\omega)$ such that $\nu\{H\} = \int_H h(\omega)\mu\{d\omega\}$ for any $H \in S$. In [72] proved a theorem based on the results of [70]:

Let $(S, \mathcal{B})$ be an arbitrary measurable space, $\Lambda$ is the set of all probability measures on $(S, \mathcal{B})$. If $\Pi \subset \Lambda$ is defined by

$$\Pi = \{ P : \int f_i dP \geq 0, \ i = 1,...,k \}, \tag{5}$$

where $f_1,...,f_k$ are given measurable functions on $(S, \mathcal{B})$, for a probability measure $Q \in \Lambda$ we have $D = D(\Pi\|Q) < \infty$ if there exists a $P \in \Pi$ with $P \ll Q$. Then the generalized $I$–projection $P^*$ of $Q$ on $\Pi$ has $Q$-density of form

$$\frac{dP^*}{dQ} = \begin{cases} \exp\{D + \sum_{i=1}^{k} \vartheta^*_i f_i\} & \text{on } \{s : (f_1(s),...,f_k(s)) \in M\} \\ 0 & \text{elsewhere} \end{cases}, \tag{6}$$

where $M$ is a linear subspace of $R^k$ and $\vartheta^* = (\vartheta^*_1,...,\vartheta^*_k) \in R^k_+$. The relation (6) holds with $M = R^k$ i.e., $P^*$ belongs to the exponential family $\{ P_\vartheta : \vartheta \subset \Theta \}$ defined by (compare with (3))

$$\frac{dP_\vartheta}{dQ} = \frac{\exp(\sum_{i=1}^{k} \vartheta_i f_i)}{\int \exp(\sum_{i=1}^{k} \vartheta_i f_i) dQ}, \qquad \Theta = \{\vartheta = (\vartheta_1,...,\vartheta_k): \int \exp(\sum_{i=1}^{k} \vartheta_i f_i) dQ < \infty\}, \qquad (7)$$

if there exists a $P \in \Pi$ with $P \equiv Q$, where $P \equiv Q$ designates mutual absolute continuity. Under the last condition

$$D = D(\Pi \| Q) = \max_{\vartheta \in R_+^k} [-\log \int \exp(\sum_{i=1}^{k} \vartheta_i f_i) dQ],$$

where the maximum is attained if $P_\vartheta = P^*$.

Here $R^k$ is the $k$-dimensional Euclidean space, $R_+^k$ is the half-space in $R^k$, defined by the condition $x^k > 0$. The probability measure $P^*$ will be called the $I$ – projection of $Q$ on $\Pi$ if $P^* \in \Pi$ and $D(P^* \| Q) = D(\Pi \| Q)$. In our case, probability measure $Q=z=\{q_1,p_1,...,q_N,p_N\}$ is phase space; $\vartheta_i$ $=F_i$ from (3), $f_i=P_i(=u,\ T_\gamma)$; $k=2$, $\vartheta_1=\beta$, $f_1=u$, $\vartheta_2=\gamma$, $f_2=T_\gamma$ (8). The integral $\int f_i dP > 0$ in condition (5) is the Lebesgue integral of a given measurable function $f_i$ with respect to the probability measure $P \in \Pi$ with $P \ll Q$. If the integration is carried out over the entire probability space $\Omega$, then this integral is equal to the mathematical expectation of the function $f_i$, if the function $f_i$ is non-negative. The non-negativity condition is satisfied for $f_2=T_\gamma$.

The results of this theorem can be compared with the application of the maximum entropy principle when used only the first moment as information of the density function, when a Gibbs or exponential distribution of the form (7) is considered. The extremum of the relative entropy $D(P\|Q)$, the Kullback entropy, is determined. The same operation for finding a non-equilibrium distribution density similar to (7) is performed in [52]. It is written in [72]: "Such results are relevant for statistical physics,…such conditional limit theorems provide a justification of the "maximum entropy principle" in physics".

Using (7) with $k=2$, $\vartheta_1=\beta$, $f_1=u$, $\vartheta_2=\gamma$, $f_2=T_\gamma$, for function $\rho_{rel}(z;u,T_\gamma)$ from (4) we obtain

$$\rho_{rel}(z;u,T_\gamma) = \frac{e^{-\beta u(z) - \gamma T_\gamma(z)}}{Z(\beta,\gamma)}. \qquad (8)$$

Substituting the expression (8) in the relation (4) and replacing the variables, passing from the variables $z$ to the variables $u$, $T_\gamma$, we get

$$p(u,T_\gamma) = \frac{e^{-\beta u - \gamma T_\gamma} \omega(u,T_\gamma)}{Z(\beta,\gamma)}, \qquad (9)$$

where $\beta = 1/T$ is the inverse temperature of the reservoir ($k_B = 1$, $k_B$ is Boltzmann constant),

$$Z(\beta,\gamma) = \int e^{-\beta u - \gamma T_\gamma} dz = \iint du\ dT_\gamma\ \omega(u,T_\gamma) e^{-\beta u - \gamma T_\gamma} \qquad (10)$$

is the partition function, $\beta$ and $\gamma$ are the Lagrange multipliers satisfying the following expressions for the averages:

$$\langle u \rangle = -\frac{\partial \ln Z(\beta,\gamma)}{\partial \beta}\bigg|_\gamma, \qquad \langle T_\gamma \rangle = -\frac{\partial \ln Z(\beta,\gamma)}{\partial \gamma}\bigg|_\beta. \qquad (11)$$

The factor $\omega(u)$ in the case of a distribution for $u$ is replaced by $\omega(u,T_\gamma)$, which is the volume of the hypersurface in the phase space containing fixed values of $u$ and $T_\gamma$. If $\mu(u,T_\gamma)$ is the number of states in the phase space with parameter values less than $u$ and $T_\gamma$, then $\omega(u,T_\gamma)=d^2\mu(u,T_\gamma)/dudT_\gamma$. Moreover, $\int \omega(u,T_\gamma)dT_\gamma=\omega(u)$. The number of phase points with parameters in the interval between $u$, $u+du$; $T_\gamma$, $T_\gamma+dT_\gamma$, is $\omega(u,T_\gamma)dudT_\gamma$. Similarly, as the distribution function in [53, 66] is dimensionless and normalized to the minimum size of the phase volume, which in the one-dimensional case is equal to the Planck constant $h=2\pi\hbar$, let's make the function $\omega(u,T_\gamma)$ dimensionless by multiplying it by some unit standard *FPT* $\tau_{FPT}$.

In expressions (2), (6)-(7) the distributions are given in a general way. Thus, expressions (6)-(7) are valid for arbitrary random variables that satisfy condition (5). The explicit form of these distributions for random variables $u$, $T_\gamma$ is given in (4), (8)-(11), and the explicit form *FPT* distribution of a random process of *J* for the flux level $J_{thr}$ is in (18), Section 4.

In expressions (8)-(11), the values of energy $u$ and *FPT* $T_\gamma$ are chosen as thermodynamic parameters. The value, thermodynamically conjugated to *FPT*, $\gamma$ is associated with the production and flow of entropy, which characterize non-equilibrium processes in an open statistical system. At $\gamma=0$ and $\beta=\beta_0=T^{-1}_{eq}$, where $T_{eq}$ is the equilibrium temperature, the non-equilibrium distribution (8) converges to equilibrium Gibbs distribution. The distribution (8) over $u$ and *FPT* is a generalization of Gibbs distribution to a non-equilibrium situation. The canonical Gibbs distribution is obtained from the microcanonical ensemble in the zero approximation by the interaction of the system with the environment. Using the *FPT* $T_\gamma$ (1), an effective account is taken of this interaction (similarly to the methods of McLennan [73] and Zubarev *NSO* [53, 54], and [69]). The value of $\gamma$ can be considered as a measure of deviation from equilibrium.

Let us make a remark about the form of the distribution function (8). The value appearing in (8) is $\exp\{-\gamma T_\gamma\}$ corresponds to the exponential limit distribution describing infinitely large times. At the same time, the distributions (7), (8) characterizes the guiding statistics [71]. The accuracy of this approximation, based on the theory of rare events, is estimated at [70]. Refinement of the distribution is possible, for example, by taking into account the quadratic terms in $T_\gamma$ in the exponent (8). In the proposed approach $\exp\{-\gamma T_\gamma\}$ acts as the kernel of the integral Laplace transform (16). A more detailed distribution of the value $T_\gamma$ is contained in the distribution function of *FPT* $f(T_\gamma)$ in $\omega(u,T_\gamma)$ (9), (13), which can describe finite arbitrarily small time intervals.

## 3. Generalized thermodynamics of systems with parameter of *FPT*

In the previous section, we found an expression for the *FPT* distribution. We need to find an expression for the conjugate thermodynamic quantity $\gamma$. Assuming that *FPT* can be instrumentally measured, we introduce the local specific entropy $s_\gamma$ corresponding to the distribution (8) ($u$ is specific internal energy) by the relation [53, 54, 59]

$$s = -\langle \ln \rho_{rel}(z;u,T_\gamma) \rangle = \beta\langle u \rangle + \gamma\langle T_\gamma \rangle + \ln Z(\beta,\gamma); \qquad ds = \beta d\langle u \rangle + \gamma d\langle T_\gamma \rangle. \qquad (12)$$

Expression (12) is the Legendre transform for the variables $u$ and $T_\gamma$. For spatially inhomogeneous systems, the values $\beta$ and $\gamma$ in the general case depend on the spatial coordinate. The distribution (8) can be considered for a small volume element in which the values of $\beta$ and $\gamma$ are replaced by the average values constant over this volume element. In non-equilibrium thermodynamics, the densities of extensive thermodynamic quantities (entropy, internal energy, mass fraction of a component) are considered. We follow this approach, including here the *FPT* parameter, with $T_\gamma \to T_\gamma/V$, where *V* is volume. The assumption that volume elements are small and homogeneous is in fact the hypothesis of local equilibrium adopted in classical non-equilibrium thermodynamics. However, in extended irreversible thermodynamics (*EIT*) [43], where thermodynamic flows variables close to *FPT* are used, the hypothesis of local equilibrium is not applied.

In the distributions (8)-(9), containing the *FPT* as a thermodynamic parameter, the joint probability for the quantities $u$ and $T_\gamma$ is (9). The distribution (9) corresponds to the general formulas (2), (4). The factor $\omega(u,T_\gamma)$ is the joint probability for $u$ and $T_\gamma$, considered as the stationary probability of this process. We rewrite the value $\omega(u,T_\gamma)$ in the form

$$\omega(u,T_\gamma) = \omega(u)\omega_1(u,T_\gamma) = \omega(u)\sum_{k=1}^{n} R_k f_{1k}(T_\gamma,u). \tag{13}$$

In Eq. (13) it is assumed that there are *n* classes of states in the system; $R_k$ is the probability that the system is in the *k*-th class of states, $f_{1k}(T_\gamma,u)$ is the density of the distribution of *FPT* $T_\gamma$ in this class of (ergodic) states (in the general case, $f_{1k}(T_\gamma,u)$ depends on *u*). As a physical example of such a situation (characteristic of metals, glasses, etc.), a potential of many complex physical systems can be mentioned. In what follows we consider the case *n=1* only. Such situation is set up in [74-76]. The minimum points of the potential correspond to metastable phases, disordered structures, etc. The phase space of these systems is divided into isolated regions, each of which corresponds to a metastable thermodynamic state, and the number of regions increases exponentially with increasing total number of (quasi)particles [74].

The form of the function $f_{1k}(T_\gamma,u)$ in (13) reflects not only the internal properties of the system, but also the impact of the environment on the open system, and its interaction with the environment. A physical interpretation of the density of the exponential distribution in *NSO* for the function $f_{1k}(T_\gamma,u)$

$$f_{1k}(T_\gamma,u) = T_0^{-1} exp\{-y/T_0\}, \qquad T_0 = T_{\gamma|\gamma=0}, \tag{14}$$

is given in [59]: the system evolves as an isolated system controlled by the Liouville operator. In addition, the system undergoes random transitions, and the phase point representing the system switches from one trajectory to another with an exponential probability under the influence of a "thermostat". Exponential distribution would describe a completely random system. The impact of the environment on the system can also be included, for example, this applies to systems in a non-equilibrium state with unsteady flows at the input and output. The nature of the interaction with the environment may be different; therefore, various forms of the function $f_{1k}(T_\gamma,u)$ can be used [42]. The exponential distribution is the limiting distribution that is valid in the limit of infinitely large times. Since *FPT*s are finite, in the general case their behavior is characterized by a variety of distributions, the form of which depends on the interaction of the system with the environment and the random processes governing the system itself.

Note that a value similar to $\gamma$ is determined in [77]–[79] for a fractal repeller object. It is equal to zero for a closed system, and for an open system it is equal to $\Sigma\lambda_i$-$\lambda_{KS}$, where $\lambda_i$ are Lyapunov exponents, $\lambda_{KS}$ is the Kolmogorov-Sinai entropy. The initial state is not necessarily an equilibrium, but a stationary non-equilibrium state. For this case, we can write relations similar to (8) – (11), (12). Deviations from the equilibrium state are considered below.

It is shown in [40-41] that the parameter $T_0$ of the exponential distribution (14) of the *FPT*, which is equal to the average unperturbed *FPT*, does not depend on a random value of energy, but rather depends on the average value of equilibrium energy and inverse equilibrium temperature. In this case, the distribution $f_{1k}(T_\gamma,u)$ (14) is independent of random energy *u*, the integration variables are factorized, and the partition function (10) is written as the product of the equilibrium and non-equilibrium factors, $Z(\beta,\gamma)=Z_\beta Z_\gamma$. Let us assume that the distribution $f_{1k}(T_\gamma,u)$ is independent of the random variable *u* and this independence holds for other kinds of distributions different from the form (14). The non-equilibrium part of the partition function in this case is the Laplace transform of the distribution of the *FPT*. The average *FPT* and higher moments depends on the non-equilibrium parameter $\gamma$, which is expressed in terms of the difference between the equilibrium and non-equilibrium entropy from Eq. (16), (17). When the non-equilibrium parameter $\gamma$ and the entropy difference tends to zero, the average *FPT* of the system assumes an equilibrium value $T_0$, which also supposed to be finite due to the open character of the system and the presence of fluctuations.

$$Z(\beta,\gamma)=Z_\beta Z_\gamma, \quad Z_\beta = \int e^{-\beta u}\omega(u)du, \quad Z_\gamma = \int_0^\infty e^{-\gamma T_\gamma}\sum_{j=0}^n P_j f_j(T_\gamma,u)dT_\gamma \quad Z_\gamma = \int_0^\infty e^{-\gamma T_\gamma} f(T_\gamma)dT_\gamma, \tag{15}$$

where $f(T_\gamma)$ is the probability density of the distribution of the *FPT*.

The thermodynamic relations for the entropy (12) are written in [40-41]. We rewrite (12) in the form of the relationship of expressions for equilibrium and non-equilibrium entropy. The parameter $\gamma$ of distribution (1) will be related to entropy using the generally accepted definition of entropy in statistical physics as the logarithm of the distribution density (8) averaged over this distribution, $s = \langle \ln \rho(z,u,T_\gamma) \rangle$, where brackets denote averaging. Variables are separated as in (15). In the case of one class of ergodic states, from (13) we obtain $\omega(u,T_\gamma) = \omega(u) f(T_\gamma)$, $\omega(T_\gamma) = f(T_\gamma)$. The distribution density (9) is equal to $p(u,T_\gamma) = \dfrac{e^{-\beta u} \omega(u)}{Z_\beta} \dfrac{e^{-\gamma T_\gamma} f(T_\gamma)}{Z_\gamma} = p(u) p(T_\gamma)$, where $Z_\beta$, $Z_\gamma$ defined in (15). The entropy of this distribution is $s = -k_B \int p(u,T_\gamma) \ln[p(u,T_\gamma)] du dT_\gamma = s_\beta + s_\gamma$, $s_\beta = \beta \bar{u} + \ln Z_\beta$, $s_\gamma = \gamma \bar{T}_\gamma + \ln Z_\gamma$. We took into account that for the nonequilibrium case in [52], the relation $s = s_B + \langle s(B) \rangle$, $s_B = -k_B \int p(B) \ln[p(B)] dB$, $\langle s(B) \rangle = k_B \int p(B) \ln[\omega(B)] dB$ was obtained; $B$ are random internal thermodynamic parameters, functions of dynamic variables $z$; in our case $B_1 = u$, $B_2 = T_\gamma$. The total uncertainty in the system is equal to the sum of the uncertainty of the parameters $B$ and the average uncertainty of the dynamic variables remaining after fixing the parameter $B$. Expanding in a power series in powers $\gamma$ value $s_\gamma$, we get $s_\gamma = -\gamma^2(\langle T_0^2 \rangle - \langle T_0 \rangle^2) \le 0$, $s \to s/k_B$, entropy is divided by $k_B$, Boltzmann's constant.

Thus, the terms $\int \dfrac{e^{-\beta u} \omega(u)}{Z_\beta} \ln[\omega(u)] du$, $\int \dfrac{e^{-\gamma T_\gamma} f(T_\gamma)}{Z_\gamma} \ln[f(T_\gamma)] dT_\gamma$ cancel and

$$s = s_\gamma + s_\beta = s_\beta - \Delta = \gamma \bar{T}_\gamma + \beta \bar{u} + \ln Z = \beta \bar{u} + \ln Z_\beta - \Delta, \quad -\Delta = s_\gamma = s - s_\beta, \quad -\Delta = s_\gamma = \gamma \bar{T}_\gamma + \ln Z_\gamma, \quad (16)$$

where the quantities $Z_\gamma$, $\bar{T}_\gamma$, are defined in the relations (10)-(11), (15), $\Delta \ge 0$; $Z_\beta = \int e^{-\beta u} \omega(u) du$, $Z_\beta$ is partition function; $\bar{u}$ is energy; $s_\beta$ is entropy related to the parameter $u$, ($s$ is the entropy density). We got a match with the expression $s = \langle \ln \rho(z,u,T_\gamma) \rangle$.

For the extended irreversible thermodynamics (*EIT*) [43] and for the case of thermal conductivity $\Delta = a_\beta q^2 / 2$, $a_\beta = \tau / \rho_v \lambda \theta^2$, where $\theta^{-1}$ is the non-equilibrium temperature, $\rho_v$ is the mass density, $\lambda$ is the thermal conductivity, $q$ is the heat flux, $\tau = \tau_q$ is the correlation time of the fluxes from the Maxwell-Cattaneo equation of the form $q = -\lambda \nabla T - \tau_q \partial q / \partial t$.

We also note that the quantity $Z_\gamma = \int_0^\infty e^{-\gamma T_\gamma} f(T_\gamma) dT_\gamma$ (15) has probabilistic meaning: $Z_\gamma = \int_0^\infty e^{-\gamma T_\gamma} f(T_\gamma) dT_\gamma = P\{T_\gamma \le \tau\}$; $P\{\tau > t\} = \exp\{-\gamma t\}$. At $\gamma \sim \sigma_s$, $\sigma_s$ is entropy production, $P\{\tau > t\} \sim 1$, when $\sigma_s \to 0$, and $P\{\tau > t\} \sim 0$, when $\sigma_s \to \infty$; $P\{T_\gamma \le \tau\} \sim 1$, $\tau \to 0$.

In relation (15), the function $Z_\gamma$ (16) is assumed to be known. From the expressions (11), (15), the values $\langle T_\gamma \rangle$ are determined. They are substituted into the relations (16), (17) which is considered as an equation for the value $\gamma$. The second Lagrange multiplier $\gamma$ is determined not from the maximum entropy Ansatz, but from the algebraic equation (16), as a function of the deviation of entropy from the equilibrium value $\Delta$. In equation (16), for the value $\gamma(\Delta)$, we define the value $\Delta$ in accordance with the nature of random process. Setting a different value of $\Delta$ will change the values of $\gamma$, $T_\gamma$, $\langle T_\gamma \rangle$. By changing or including additional thermodynamic forces, we change the material flows in the system, *FPT* $T_\gamma$, $\langle T_\gamma \rangle$.

The dependence of *FPT* on the phase coordinates is important, which makes it possible to determine *FPT* as a thermodynamic parameter [51, 52]. Information about dynamic properties of the system is also required. We need to know the Laplace transform of the *FPT* distribution. This value is included in the non-equilibrium part of the partition function. In the proposed approach, the Laplace transform parameter of the *FPT* distribution coincides with the thermodynamic parameter $\gamma$, which is the conjugate to the random thermodynamic parameter *FPT* $T_\gamma$. The factor *exp{-$\gamma T_\gamma$}* from the statistical distribution (8) for energy and *FPT* is the kernel of the integral Laplace transform (15). The relationship (12), (16) for the deviation of entropy from its equilibrium value includes such quantities as the product of the average *FPT* $\langle T_\gamma \rangle$ and its conjugate thermodynamic parameter $\gamma$, $\gamma \langle T_\gamma \rangle$. The average value of *FPT* $\langle T_\gamma \rangle$ depends on the parameter $\gamma$. From the equation for entropy (16), this parameter is determined, which is then substituted into the expressions for $\langle T_\gamma \rangle$ and in the partition function. The obtained results are general and thus valid for various systems of any nature. The physical approach is based on a general mathematical definition of stop times. This allows to control the speed and time needed for a stochastic process to achieve an arbitrary level. Section 4 and Appendix A provide examples of entropy changes affecting the average *FPT*.

## 4. Examples. *FPT* for flow values, for level *L* in drift-diffusion process

In this work the theory does not allow to determine the first passage statistics itself. The "incorporation" of first passage statistics would naturally arise as a choice of the random process that describes the phenomenon under study and the appropriate *FPT*. There are many papers exploring FPT statistics [1-7, 10-35, 37-39, 47, 56-57, 81, 84, 97-99]. A part of them considers general mathematical aspects only [1-6, 10-12, 15, 23-35, 47, 56-57, 81], others refer to a variety of applications [4-7, 13-14, 16-22, 37-39, 84, 97-99] with numerous types of FPT statistics relevant to the specific problem. Any of them can be used to illustrate the approach developed in this article.

The main problem is to establish an adequate correspondence between a physical phenomenon and the stochastic process used for its mathematical description. As to the *FPT* problem, the variety of existing stochastic processes is available and can be adapted for a specific problem. In each of such processes, changes in the boundary conditions can significantly affect the properties of the process (for example [4, 22]). It is possible to single out, for example, random processes where the average *FPT* can only decrease under various impacts on the system. However, a class of random processes can be envisaged where certain external impacts would increase the average *FPT*, although generally they are more sophisticated for the analytical treatment.

What are criteria for selecting a random process and the corresponding *FPT*? The ultimate set would be: an optimal match of the random process to the system; at least qualitative agreement with the experimental results; some inner criteria. In general, this is the subject of a separate study. The behavior of the same phenomenon under different conditions can be different and can be described by different random processes, distributions and their Laplace transforms. Examples of this kind are given in [83]. We are not much interested in *FPT* distributions themselves, but it is their Laplace transforms what matters. This facilitates the analytical treatment.

*FPT* statistics is determined by the random process that characterizes the phenomenon under study. In cases where an exact solution cannot be found, one can use general results for *FPT*, for example, the first few terms of the expansions obtained in [27-28] for the Laplace transform of *FPT*, which determine the quantities included in them depending on the process under consideration. There are other options for selecting and setting up *FPT* statistics models. The use of various kinds of approximations also substantially depends on the stage of evolution of the system [32, 39].

The range of applicability of the results is extensive and, apparently, corresponds to the range of applicability of Gibbs statistics. This generality allows to include arbitrary distributions rather than restricting to any one specific *FPT* forms.

Let us illustrate the application of the above approach by examples. In [16] and [17] *FPT*s are considered for achieving a certain level of entropy production and flux values. The above possibilities of connecting the *FPT* distribution with the thermodynamic characteristics of the system are applicable to this situation, as well as to others cases. Consider reaching a certain value of the flow level. As an independent thermodynamic parameter, we choose a random time mark $T_{\gamma J}$ of the first achievement of a given flow value $J_{thr}$. Adding this parameter, along with the energy, to the exponential distribution, as in (8), we write the relation for the entropy of the form (12), (15) taking into account (16)

$$s_{\gamma J} = \beta \bar{u} + \gamma \bar{T}_{\gamma J} + \ln Z = \beta \bar{u} + \ln Z_\beta - \Delta, \qquad -\Delta = \gamma \bar{T}_{\gamma J} + \ln Z_\gamma, \qquad \Delta = s_\beta - s_{\gamma J}. \qquad (17)$$

As in [17], for the *FPT* distribution density of the flux level $J_{thr}$, we use the inverse Gaussian distribution (Wald's distribution) with a density of the form

$$F_{bnd}(t) = \sqrt{\frac{J_{thr}\sigma\langle t\rangle^2}{4\pi t^3}} \exp[-\frac{J_{thr}\sigma(t-\langle t\rangle)^2}{4t}], \qquad (18)$$

where $\sigma$ is entropy production rate (with Boltzmann's constant $k_B = 1$), $\langle t \rangle$ is average (with $\gamma=0$) value of *FPT*. The quantity $X$ from (1) in this case is equal to the flow $J$. The Laplace transform of distribution (18) is

$$Z_{J\gamma} = \int_0^\infty e^{-\gamma t} F_{bnd}(t)dt = \exp\{\frac{aT_{J0}}{2}(1-\sqrt{1+4\gamma/a})\}, \qquad a = J_{thr}\sigma, \qquad (19)$$

where $\langle t\rangle = \bar{T}_{J0} = T_{J0}$ is the value of $\bar{T}_{\gamma J}$ in the absence of disturbances, at $\gamma = 0$. The expression obtained from relation (19) for the average time to reach the level $J_{thr}$ is

$$\bar{T}_{\gamma J} = -\frac{\partial \ln Z_{J\gamma}}{\partial \gamma}\bigg|_\beta = \frac{T_{J0}}{\sqrt{1+4\gamma/a}}, \qquad T_{J0} = \langle t \rangle. \qquad (20)$$

Substituting (19), (20) into (17), we obtain, taking into account (16), a quadratic equation to determine the dependence $\gamma(\Delta)$ with the solution

$$\gamma = \frac{a}{2}[(1+\frac{2\Delta}{aT_0})(1\pm\sqrt{(1+\frac{2\Delta}{aT_0})^2-1}))-1]. \qquad (21)$$

In this case, there are two branches of the solution with the condition $\gamma_{|\Delta=0} = 0$.

The value $\Delta$ depends on the type of flows $J$. For example, for heat fluxes in *EIT* [43] $J = q$, $\Delta = a_\beta q^2/2$, $a_\beta = \tau/\rho\lambda\theta^2$.

Substitution of (21) into (20) gives the relation

$$\bar{T}_{\gamma J} = \frac{T_{J0}}{\sqrt{1+2[(1+\frac{2\Delta}{aT_0})(1\pm\sqrt{(1+\frac{2\Delta}{aT_0})^2-1}))-1]}} \ .$$

Knowledge of the value $\gamma$ (21) also makes it possible to use the probabilistic interpretation of the Laplace transform $Z_{J\gamma} = \int e^{-\gamma T_{\gamma J}} P(T_{\gamma J})dT_{\gamma J} = P(T_{\gamma J} \le \tau), \quad P(\tau > t) = e^{-\gamma t}$.

The same distribution (18) [80] is written in [22] and is applied in [16] to various problems, including for *FPT* level *L* in drift-diffusion process. The Langevin equation for drift-diffusion process is

$$\frac{dX(t)}{dt} = v + \zeta(t),$$

where *X(t)* from (1) is the coordinate, $v = F/\varsigma$ is the drift velocity, *F* is an external force, $\varsigma$ is a friction coefficient, and $\zeta(t)$ is a Gaussian white noise with zero mean $\langle\zeta(t)\rangle = 0$ and with autocorrelation $\langle\zeta(t)\zeta(t')\rangle = D\delta(t-t')$, $D=k_BT/\varsigma$ is the diffusion coefficient. In [16] the *FPT* distribution for *X* to pass at time $T_\gamma = t$ for the *FPT* the threshold *L*>0, starting from the initial condition *X(0) = 0*, is given by Wald's distribution [80, 16]

$$F_{T_\gamma}(t) = \sqrt{\frac{|L|^2}{4\pi Dt^3}} \exp[-\frac{(L-vt)^2}{4Dt}]. \tag{22}$$

Distributions (18) and (22) coincide for $\langle t \rangle = L/v$, $a = v^2/D$ in (22) (in (18)-(19) $a = J_{thr}\sigma$). For the deviation of entropy from the equilibrium value during diffusion in the time $\overline{T}_\gamma$ can write the expression from *EIT* [43]

$$\Delta = s_\beta - s = \dot{s}\overline{T}_\gamma/\rho, \quad \dot{s} = \sum_{k=1}^{N}\frac{\mu_k}{T}\nabla J_k, \quad J_k = \rho_k(v_k - v), \tag{23}$$

$\mu_k$ is the chemical potential. Expression for $\overline{T}_\gamma$ we write from (19) - (20), (22). From the ratio

$$\gamma\overline{T}_\gamma + \ln Z_\gamma = -\Delta = -\dot{s}\overline{T}_\gamma/\rho \tag{24}$$

we obtain a quadratic equation for *γ* with the solution

$$\gamma = \frac{a}{4}[1 + \frac{\dot{s}}{\rho}\frac{a}{4} - 2\sqrt{1 + \frac{\dot{s}}{\rho}\frac{a}{4}}]. \tag{25}$$

We assume that $u_\gamma = -\partial \ln Z_\gamma/\partial\beta|_{\gamma=0} = 0$. Derivatives of *v* and *a* are equal $\frac{\partial T_0}{\partial\beta} = -\frac{T_0}{v}\frac{\partial v}{\partial\beta}$, $\frac{\partial a}{\partial\beta} = a(\frac{2}{v}\frac{\partial v}{\partial\beta} - \frac{1}{D}\frac{\partial D}{\partial\beta})$, $\frac{\partial D}{\partial\beta} \approx -E_a D$, $E_a$ is the activation energy. From the condition $u_\gamma = 0$ we find

$$\frac{1}{v}\frac{\partial v}{\partial\beta} = -E_a\frac{\sqrt{1+4\gamma/a} - 1 - 2\gamma/a}{\sqrt{1+4\gamma/a} - 1}. \tag{26}$$

By $\gamma \to 0$, $\partial v/\partial\beta \to 0$;

$$\frac{1}{T_0}\frac{\partial T_0}{\partial\beta} = E_a\frac{\sqrt{1+4\gamma/a} - 1 - 2\gamma/a}{\sqrt{1+4\gamma/a} - 1}, \quad \frac{1}{a}\frac{\partial a}{\partial\beta} = -E_a\frac{\sqrt{1+4\gamma/a} - 1 - 4\gamma/a}{\sqrt{1+4\gamma/a} - 1}. \tag{27}$$

For small γ, $\frac{1}{T_0}\frac{\partial T_0}{\partial\beta} \approx -E_a\frac{2\gamma}{a}$, $\gamma^2 \approx \frac{\dot{s}}{\rho}\frac{a}{2\beta E_a}$. Substitution of (25) into (20) gives the relation

$$\overline{T}_\gamma = \frac{T_0}{\sqrt{2 + \frac{\dot{s}}{\rho}\frac{4}{a} - 2\sqrt{1 + \frac{\dot{s}}{\rho}\frac{4}{a}}}}. \tag{28}$$

For molecular diffusion in perfect fluid mixtures in [43] the expressions for the Helmgoltz free energy $\Psi = u - Ts$ were obtained $\Psi_{eq} - \Psi = -\frac{1}{2}\frac{\tau L}{\rho}JJ = -K$, where $J=J_1$ is the diffusion flux of the first component, $\tau$ is the relaxation time of *J*. Using expressions (16), (17), we find

$$\Psi_{eq} - \Psi = -T(s_{eq} - s) = -T\Delta = -K, \quad \gamma \bar{T}_\gamma + \ln Z_\gamma + \beta K = 0 \quad (17).$$ Solving this equation for γ, we find

$$\gamma = R[\frac{2R}{aT_0^2} + \frac{1}{T_0}\sqrt{1 + \frac{4R^2}{(aT_0)^2}}], \quad R = \sqrt{K(K + aT_0)}, \quad \bar{T}_\gamma = T_0 / \sqrt{1 + \frac{4R}{aT_0}[\frac{2R}{aT_0} + \sqrt{1 + \frac{4R^2}{(aT_0)^2}}]} \quad . \quad (29)$$

Relation (29) can be rewritten as

$$\bar{T}_\gamma = \frac{T_0}{\sqrt{1 + 2p(p + \sqrt{1 + p^2})}}, \quad D = D_0\sqrt{1 + 2p(p + \sqrt{1 + p^2})}, \quad p = 2R/aT_0. \quad (30)$$

Appendix A shows that it is possible to control the Feller process to increase mean *FPT*.

## 5. Conclusion

Now we briefly formulate the significance and possible outcome of this work.

First, the random variable *FPT* is considered as a thermodynamic parameter. A generalized Gibbs distribution is introduced containing *FPT* as a thermodynamic parameter. This distribution of the form (8) - (9) is important in many areas. For example, in [87] from a distribution of the form (9) containing *FPT* in the form lifetime, the distribution of neutron energy in a nuclear reactor was obtained. This theoretical distribution coincides with the experimentally observed energy distribution of neutrons in a nuclear reactor: Maxwell distribution for thermal neutrons and Fermi distribution for fast neutrons with a transition region between them. The use of a distribution of the form (9) is justified by the finite lifetime of neutrons, due to which the equilibrium state of the system of neutrons with moderator nuclei is not achieved. The finite lifetime and size are characteristic of many systems: atomic nuclei, liquid droplets [88], micelles [89], formation in biological membranes [90-92], etc.

Second, the new physical interpretation of *FPT* makes it possible to relate *FPT* to thermodynamic quantities, in particular, to the change in the entropy of the system. Knowledge of such patterns allows to control the processes occurring in the system. In many cases, the pace of events is important and often decisive. The events themselves refer to the moments of stopping, of which *FPT* is a special case. The change in entropy includes various kinds of effects on the system, which can speed up or slow down the development of the system. Generally, it is not known, how to control the parameters that enter first passage statistics explicitly. In some cases, it might be easier to control first passage times by adjusting entropy changes. Much depends on the specific physical situation. For example, in the exponential distribution (18), a controllable parameter can be the average *FPT* $T_0$. Then, the approach of the present paper can be applied. For the distribution in the form of Eq. (14) this was done in [40, 41]. Thus, it is possible to investigate the possibility of controlling the rate of micelle formation [89], formation of aerosols [93], evaporation and condensation of droplets [88], nucleation of a new phase, etc. In [89-93] a stochastic storage model was deployed for the description of the processes of micelle formation and the behavior of aerosols and lipid rafts [special areas (microdomains) of the plasma membrane, enriched with glycosphingolipids and cholesterol] in biological membranes. Accordingly, *FPT* for these systems can be obtained from stochastic storage models.

The third important point is to obtain numerous connections between the stochastic theory *FPT* - and these theories are diverse, including the theory of queues, the stochastic storage theory, etc. - and statistical physics and non-equilibrium thermodynamics. Such connections will create many applications, both among those where *FPT* has already been used, but at a new level, taking into account the obtained patterns, and in new directions. For example, application to the Kramers problem and related applications, to the behavior of various physicochemical systems (emulsions, foams, aerosols, micelles, etc.).

General relationships are important, the implementation of which is difficult to predict. Generalizing ideas, if correct, should lead to a large number of specific applications. The results obtained in this work are also more general, perhaps the most important. Since *FPT* characterizes the behavior of a system in time, it can be hoped that, perhaps, further studies of the connections between *FPT* and non-equilibrium statistical physics and non-equilibrium thermodynamics will make it possible to advance in understanding such still largely mysterious phenomena, such as the lifetime and time in general.

This paper discussed the change in the entropy of a non-equilibrium statistical system during the processes occurring in it, when a random process describing its certain physical quantity reaches some given boundary, is associated with the Laplace transform parameter of the *FPT* distribution and with the thermodynamic parameter conjugate to *FPT*. The Laplace transform of the *FPT* distribution and its derivatives, *FPT* moments are expressed in terms of the change in the entropy and other thermodynamic quantities. This is achieved by choosing *FPT* as an additional thermodynamic parameter. Besides (18), (22), (31), (33) many other *FPT* distributions for various processes [27-29] and physical, biological and other systems can be considered [1-3].

The proposed approach combines mathematical *FPT* research with non-equilibrium statistical physics and statistical thermodynamics. In this case, a generalization of the well-known Gibbs ensemble is used. This allows us to consider a large number of problems related to both the Gibbs distribution and the set of *FPT* problems noted in the introduction. However, knowledge of the mathematical results of the *FPT* of the stochastic study is necessary to determine the Laplace transform of the *FPT* distribution. A new look opens up new possibilities for both various aspects of statistical thermodynamics and conventional first passage theory. Practical applications of the proposed approach are wide and varied. This is both obtaining different distributions for many phenomena associated with superstatistics [8, 9, 103], and non-equilibrium thermodynamics [94, 95], and the behavior of neutrons in nuclear reactors [87, 96], and the behavior of aerosols [93] and rafts in biological membranes [92-94] (storage models were used for them, but the approach of this work is also applicable). The proposed method is applicable, for example, to radiation-enhanced diffusion, where this method appears in integral form, without concentrating on the details of the description, to the distribution of the first-passage times for the net number of electrons transferred between two metallic islands in Coulomb blockade regime [84], to the theory of nuclear reactors, etc.

The results for *FPT* of various stochastic processes, for different values and different limits, different levels of their achievement, can be described by the proposed algorithm. The corresponding *FPT* is selected as an additional thermodynamic variable. The conjugate thermodynamic force associated with the change in entropy in the process of reaching various limits by a random process is determined.

The relevant distribution (8) obtained in Section 2 is similar to the relevant *NSO* distribution (3). However, distribution (3) does not contain the thermodynamic variable *FPT*. Distribution (8) is written as a special case of distribution (7) obtained using the results of mathematical statistics.

For exponential distribution (14), arbitrary impact on the mean *FPT*, expressed in terms of the deviation of entropy from the equilibrium value $\Delta = s_\beta - s$ (17), can only decrease the mean *FPT*. This also applies to mean *FPT* (20), at $\gamma>0$. However, there are distributions in which external impacts can increase the average *FPT*. These are, for example, the distribution (31), the Weibull distribution, which is valid in the limiting case for a two-dimensional continuous-time dynamical system, with an attracting fixed point [24], and many other distributions. In general, for the growth of the mean *FPT* $\bar{T}_\gamma$ the condition must be met $\bar{T}_\gamma > \bar{T}_{\gamma 0}$, $\bar{T}_\gamma = -\dfrac{\partial \ln Z_\gamma}{\partial \gamma}\bigg|_\beta$,

$$Z_\gamma = \int_0^\infty e^{-\gamma x} f(x)dx, \quad \int_0^\infty xf(x)dx = \bar{T}_{\gamma 0}, \quad \int_0^\infty xe^{-\gamma x}f(x)dx > \int_0^\infty e^{-\gamma x}f(x)dx \int_0^\infty xf(x)dx, \quad Z_\gamma < e^{-\gamma \bar{T}_{\gamma 0}},$$

where *f(x)* is the probability density of the distribution of the *FPT*.

The paper provides examples of the application of this algorithm. *FPT* of a certain flow level value, *FPT* level $L$ in drift-diffusion process and *FPT* Feller process are considered. Explicit dependences of the parameter $\gamma$, which is the thermodynamically conjugated of *FPT*, through changes of the entropy are obtained. For distribution (31), an analytical solution is possible only for special cases, in contrast to distributions (18), (22). The proposed approach makes it possible to determine the intervals of entropy deviation from the equilibrium value $\Delta=\Delta s$, for which the average *FPT* increases, as in (45)-(46).

Let us indicate some open questions, limitations and approximations.

- It is not determined how far from equilibrium the proposed approach is fair. One of the possibilities for estimating the accuracy of expression (7) is to consider the approximations made in obtaining (7) in [70, 72] for each individual case.

- The limitations of the proposed approach are the assumption that the distribution for *FPT* is independent of the random value of the internal energy; the need-to-know $Z_\gamma$, the Laplace transform of *FPT*.

- Chaotic regimes, conditionally Lévy processes, are not considered.

- An important role in equations (12), (15) for determining the parameter $\gamma$ is played by the quantity $\Delta$, the deviation of entropy from its equilibrium value. In the general case, this value is unknown. Various approximations are used to determine it.

- Not all first passage processes (more precisely trajectories of said process) can be reduced to a one-dimensional support. For example, bivariate Wiener process [97, 98]; in [39, 99] and in other papers was considered the first passage times which not correspond to a one-dimensional (deriving from a projection in full phase space) process. In this work, only one-dimensional processes $X(t)$ are considered.

## Appendix A: Feller process. Possibility of increasing the average *FPT*

The described approach is applicable to multiple problems with *FPT* concept. For example, in [81], using *FPT*s, problems such as the thermal motion of a small tracer in a viscous medium, adhesion bond dissociation under mechanical stress, algorithmic trading, first crossing of a moving boundary by Brownian motion, quadratic double-well potential are considered. In [81] the random variable $\tau=\inf\{t>a: X(t)>L\}$ was considered, describing the first exit time of a stochastic process $X(t)$ outside from the interval $[-L, L]$ with starting point $x_0$ at $t_0=0$. The expression for the Laplace transform of the *FPT* distribution used in [81] is more complicated than (19) and has the form

$$Z_\gamma(r_0,\gamma) = \frac{U(\frac{\gamma L^2}{4\kappa D}, \frac{d}{2}, \kappa \frac{r_0^2}{L^2})}{U(\frac{\gamma L^2}{4\kappa D}, \frac{d}{2}, \kappa)} \frac{U(0, \frac{d}{2}, \kappa)}{U(0, \frac{d}{2}, \kappa \frac{r_0^2}{L^2})} \tag{31}$$

(The same expression (31) was obtained in [25] for Feller processes), where

$$U(a,b;z) = \frac{\Gamma(1-b)}{\Gamma(a-b+1)} M(a,b;z) + \frac{\Gamma(b-1)}{\Gamma(a)} z^{1-b} M(a-b+1, 2-b; z) \tag{32}$$

is the confluent hypergeometric function of the second kind [82] (also known as Tricomi function), $M(a,b;z) = {}_1F_1(a,b,z) = \sum_{n=0}^{\infty} \frac{a^{(n)} z^n}{b^{(n)} n!}$ is Kummer function, the confluent hypergeometric function of the first kind [82] with $a^{(0)}=1$, and $a^{(n)} = a(a+1)...(a+n-1) = \Gamma(a+n)/\Gamma(a)$, where $\Gamma(z)$ is the gamma function. In (31) $\kappa = kL^2/2k_B T$. In [81] a diffusing particle of mass $m$ trapped by a harmonic potential of strength $k$ and pulled by a constant force $F_0$ is considered; the force in the Langevin equation $F(X(t))=-kX(t)+F_0$ includes the externally applied Hooke term and constant force $F_0$, $D=k_B T/\varsigma$ is the diffusion coefficient, $\varsigma$ is the drag constant.

An equation of the form (17) for $\gamma(\Delta)$ and Laplace transform (31) takes on a more complex form and it is generally difficult to find an analytical solution. But in some approximations [81] a closed analytical expression to the problem can be found.

Consider the Feller process, which is a diffusion process with linear drift and linear diffusion coefficient vanishing at zero point of coordinates. In [25], an expression was obtained for the Laplace transform of the probability density of the the Feller process $Y(t)$ to reach the boundary $x = x_c$ for $x > x_c$ for the first time ($x$ is the initial value of the process at $t=0$) in the form

$$Z_\gamma = \frac{U(\gamma, b; x)}{U(\gamma, b; x_c)} \frac{U(0, b; x_c)}{U(0, b; x)}. \tag{33}$$

This expression (33) differs from (31) only in the values of the parameters. In (33), expression (32) with parameters $a=\gamma$, $b=2\beta_1/k^2$, where $\beta_1$, $k>0$ are drift and diffusion parameters of the Feller process. The time evolution of the process is thus governed by

$$dY(t) = [-\alpha Y(t) + \beta_1]dt + k\sqrt{Y(t)}dW(t),$$

where $\alpha > 0$, $\beta_1$, $k$ are constant parameters, $W(t)$ is the Wiener process. In the approximation of small values of $a = \gamma$ in (32), (33), one can use small $\gamma$ expansion of the Kummer function [25]

$$U(a,b,x) = 1 + aU_1(x) + O(a^2), \quad U_1(x) = -\Psi(1-b) - \int_0^x U(1,1+b,z)dz, \quad \Psi(x) = \frac{\Gamma'(x)}{\Gamma(x)}, \quad 0 < b < 1.$$

In this approximation at $x_c = 0$

$$Z_{1\gamma} = \frac{1 + \gamma U_1(x)}{1 + \gamma U_1(0)}, \quad \overline{T}_\gamma = \frac{U_1(0)}{1 + \gamma U_1(0)} - \frac{U_1(x)}{1 + \gamma U_1(x)}, \quad T_0 = U_1(0) - U_1(x) = \int_0^x U(1,1+b,z)dz. \tag{34}$$

An increase of the mean *FPT*, when $\overline{T}_\gamma > T_0$, in this case is possible at

$$U_1(0) < T_0 < 2U_1(0), \quad U_1(0) = -\Psi(1-b) > 0, \quad 0 < b < 1. \tag{35}$$

We assume that $\gamma > 0$. Consider two cases. In the first case $1 + \gamma U_1(x) = 1 - \gamma(T_0 - U_1(0)) > 0$, $T_0 - U_1(0) > 0$, $U_1(x) = U_1(0) - T_0 < 0$. Then the values of the parameter $\gamma$ for which the inequality $\overline{T}_\gamma > T_0$ for the quantities from (34) is satisfied lie in the interval

$$\frac{[2U_1(0) - T_0]}{U_1(0)[T_0 - U_1(0)]} < \gamma < \frac{1}{[T_0 - U_1(0)]}. \tag{36}$$

For the case $1 + \gamma U_1(x) = 1 - \gamma(T_0 - U_1(0)) < 0$, the opposite inequality $[2U_1(0) - T_0]/U_1(0)[T_0 - U_1(0)] > \gamma > 1/[T_0 - U_1(0)]$ should hold. However, it is not the case since from (44) follows that $0 < 2 - T_0/U_1(0) < 1$.

The fulfillment of inequality (36) depends on the parameters $b = 2\beta_1/k^2$ and $x$, the initial value. Equation (17) for determining the dependence of $\gamma$ on $\Delta$ taking into account the term of the form (10), (11), (16) in this case takes the form (here the derivatives with respect to $\beta$ are written, which are defined below)

$$-\Delta = \ln(1 + \gamma U_1(x)) - \ln(1 + \gamma U_1(0)) + \frac{\gamma[U_1(0) + \beta \partial U_1(0)/\partial \beta]}{1 + \gamma U_1(0)} - \frac{\gamma[U_1(x) + \beta \partial U_1(x)/\partial \beta]}{1 + \gamma U_1(x)}.$$

Expanding the right-hand side of this equation in a series in $\gamma$, and taking into account the terms up to $\gamma^2$, we obtain the quadratic equation

$$\gamma^2 a_2 + \gamma b_2 - \Delta = 0, \quad a_2 = (U_1^2(0) - U_1^2(x))/2 + U_1(0)\beta \partial U_1(0)/\partial \beta - U_1(x)\beta \partial U_1(x)/\partial \beta, \tag{37}$$

$$a_2 = T_0[\frac{1}{2}(2U_1(0) - T_0) + \beta \frac{\partial U_1(0)}{\partial \beta}] + [U_1(0) - T_0]\beta \frac{\partial T_0}{\partial \beta},$$

$$b_2 = -\beta(\frac{\partial U_1(0)}{\partial \beta} - \frac{\partial U_1(x)}{\partial \beta}) = -\beta \frac{\partial T_0}{\partial \beta}, \quad \frac{\partial U_1(0)}{\partial \beta} = \frac{\partial U_1(0)}{\partial b}\frac{\partial b}{\partial \beta}, \quad \frac{\partial U_1(x)}{\partial \beta} = \frac{\partial U_1(x)}{\partial b}\frac{\partial b}{\partial \beta}, \quad \beta = \frac{1}{T},$$

$$b_2 = -\beta \frac{\partial b}{\partial \beta}\int_0^x \frac{\partial U(1,1+b,z)}{\partial b}dz, \quad U(1,1+b,z) = \frac{\Gamma(-b)}{\Gamma(1-b)}M(1,1+b,z) + \Gamma(b)z^{-b}e^z.$$

For the solution of the quadratic equation (37), the condition (36) takes the form

$$\frac{[2U_1(0) - T_0]}{U_1^2(0)[T_0 - U_1(0)]^2}\{T_0[2U_1(0) - T_0][\frac{1}{2}(2U_1(0) - T_0) + \beta \frac{\partial U_1(0)}{\partial \beta}] + [U_1(0) - T_0][3U_1(0) - T_0]\beta \frac{\partial T_0}{\partial \beta}\} < \Delta <$$

$$< \frac{1}{[T_0 - U_1(0)]^2}\{T_0[\frac{1}{2}(2U_1(0) - T_0) + \beta \frac{\partial U_1(0)}{\partial \beta}] + 2[U_1(0) - T_0]\beta \frac{\partial T_0}{\partial \beta}\}. \tag{38}$$

In (37)-(38) do not include explicit expressions for $\partial U_1(0)/\partial \beta$, $\partial T_0/\partial \beta$. Now let's find them. Suppose the parameter $\alpha$ in the stochastic equation for the Feller process is $\alpha = 1/T_0$. The drift parameter is $f = -\alpha Y + \beta_1$. As above, in (18) - (19), we set the parameter $f$ to be $f = L/T_0$, where $L = x - x_c$. For $x_c=0$, $L=x$, where $x$ is the initial value of the process at the initial time moment. Then $\beta_1 = f + \alpha Y = (x+Y)/T_0$. The diffusion coefficient is $D = k^2 Y$, and the parameter $b$ of (32)-(35) is equal to $b = 2\beta_2/k^2 = 2(x+Y)Y/T_0 D$. At $\partial D/\partial \beta = -E_a D$,

$$\frac{\partial b}{\partial \beta} = b(E_a - \frac{1}{T_0}\frac{\partial T_0}{\partial \beta}). \tag{39}$$

From (34) - (35) we obtain $\frac{\partial T_0}{\partial \beta} = \frac{\partial U_1(0)}{\partial \beta} - \frac{\partial U_1(x)}{\partial \beta}$, $\frac{\partial U_1(0)}{\partial \beta} = \frac{\partial \Psi(1-b)}{\partial(1-b)}\frac{\partial b}{\partial \beta}$.

From (11), (34) we find $u_\gamma = \frac{\gamma \partial U_1(0)/\partial \beta}{1 + \gamma U_1(0)} - \frac{\gamma \partial U_1(x)/\partial \beta}{1 + \gamma U_1(x)}$. As above, we assume $u_\gamma = 0$. From here

$$\frac{\partial T_0}{\partial \beta} = \frac{\partial \Psi \gamma T_0 b E_a}{1 + \gamma(U_1(0) + b\partial \Psi)}, \quad \partial \Psi = \frac{\partial \Psi(1-b)}{\partial(1-b)}, \tag{40}$$

$$\frac{\partial U_1(0)}{\partial \beta} = \frac{\partial \Psi(1 + \gamma U_1(0))bE_a}{1 + \gamma(U_1(0) + b\partial \Psi)}, \quad \frac{\partial U_1(x)}{\partial \beta} = \frac{\partial \Psi(1 + \gamma U_1(x))bE_a}{1 + \gamma(U_1(0) + b\partial \Psi)}. \tag{41}$$

Substituting (40), (41) into (37), we obtain

$$a_2 = \frac{1}{2}(U_1^2(0) - U_1^2(x)) + \frac{b\beta E_a \partial \Psi}{1 + \gamma(U_1(0) + b\partial \Psi)}[T_0 + \gamma(U_1^2(0) - U_1^2(x))],$$

$$b_2 = -\beta \frac{\partial T_0}{\partial \beta} = \frac{\gamma \partial \Psi T_0 b\beta E_a}{1 + \gamma(U_1(0) + b\partial \Psi)}.$$

Equation (37) for the parameter $\gamma$ in terms of $\Delta$ takes the form

$$\gamma^3(U_1^2(0) - U_1^2(x))[b\beta E_a \partial \Psi + (U_1(0) + b\partial \Psi)/2] + \gamma^2(U_1^2(0) - U_1^2(x))/2 - \gamma\Delta(U_1(0) + b\partial \Psi) - \Delta = 0.$$

Neglecting the terms with $\gamma^3$, we find a positive solution to the quadratic equation

$$\gamma \approx [\Delta(U_1(0) + b\partial \Psi) + \sqrt{\Delta^2(U_1(0) + b\partial \Psi)^2 + 2\Delta(U_1^2(0) - U_1^2(x))}]/(U_1^2(0) - U_1^2(x)). \tag{42}$$

At small values of $\Delta$, when

$$\sqrt{2\Delta(U_1^2(0) - U_1^2(x))}\sqrt{1 + \Delta(U_1(0) + b\partial \Psi)^2 \frac{1}{2(U_1^2(0) - U_1^2(x))}} \approx$$

$$\sqrt{2\Delta(U_1^2(0) - U_1^2(x))}[1 + \Delta(U_1(0) + b\partial \Psi)^2 \frac{1}{4(U_1^2(0) - U_1^2(x))} + ...]$$

,

Eq. (42) can be rewritten as

$$\gamma \approx \frac{1}{(U^2_1(0)-U^2_1(x))}[\sqrt{2\Delta(U^2_1(0)-U^2_1(x))} + \Delta(U_1(0)+b\partial\Psi)(1+\frac{\sqrt{2\Delta(U^2_1(0)-U^2_1(x))}(U_1(0)+b\partial\Psi)}{4(U^2_1(0)-U^2_1(x))}+...)], \quad (43)$$

or

$$\gamma = \gamma_0\sqrt{\Delta}+\gamma_1\Delta+\gamma_2\Delta^{3/2}, \quad \gamma_0 = \sqrt{\frac{2}{(U^2_1(0)-U^2_1(x))}}, \quad \gamma_1 = \frac{U_1(0)+b\partial\Psi}{(U^2_1(0)-U^2_1(x))}, \quad \gamma_2 = \gamma_0(U^2_1(0)+b^2\partial\Psi)^2/4. \quad (44)$$

We rewrite the condition $\overline{T}_\gamma > T_0$ in the form $T_0 > T_0[1+\gamma(U_1(x)+U_1(0))+\gamma^2 U_1(x)U_1(0)]$,

$$(U_1(0)+U_1(x))+\gamma U_1(0)U_1(x) < 0, \quad \gamma(-U_1(x)U_1(0)) > U_1(0)+U_1(x). \quad (45)$$

This condition can be satisfied for $U_1(x) < 0$, $U_1(x)+U_1(0) > 0$, which is possible at certain values of $x$. Substituting the expression (44) into (45) and neglecting the term with $\Delta^{3/2}$, we find the condition that $\overline{T}_\gamma > T_0$ in the form

$$\sqrt{\Delta} > \gamma_+, \quad \gamma_+ = \frac{\sqrt{(U^2_1(0)-U^2_1(x))}}{2(U_1(0)+b\partial\Psi)}[\sqrt{1-\frac{2(U_1(0)+U_1(x))(U_1(0)+b\partial\Psi)}{(U_1(0)U_1(x))}}-1], \quad (46)$$

where $\gamma_+$ is the positive solution to the quadratic equation $\gamma_1\Delta+\gamma_0\sqrt{\Delta}+\frac{U_1(0)+U_1(x)}{U_1(0)U_1(x)}=0$.

Substitution of (44) into the Eq (34) for $\overline{T}_\gamma$ leads to the following formula:

$$\overline{T}_\gamma = \frac{T_0}{R_2}, \quad R_2 = 1 + d_{1/2}\Delta^{1/2} + d_1\Delta^1 + d_{3/2}\Delta^{3/2} + d_2\Delta^2 + d_{5/2}\Delta^{5/2} + d_3\Delta^3, \quad (47)$$

$$d_{1/2} = \sqrt{\frac{2(U_1(0)+U_1(x))}{T_0}}, \quad d_1 = \frac{1}{T_0}[U_1(0)+b\partial\Psi+\frac{2}{T_0}\frac{U_1(0)U_1(x)}{(U_1(0)+U_1(x))^{3/2}}],$$

$$d_{3/2} = \sqrt{\frac{2}{T_0}}(U_1(0)+b\partial\Psi)[\frac{1}{4}\sqrt{(U_1(0)+U_1(x))}(U_1(0)+b\partial\Psi)+\frac{2}{T_0}\frac{U_1(0)U_1(x)}{(U_1(0)+U_1(x))^{3/2}}],$$

$$d_2 = U_1(0)U_1(x)\frac{\gamma_0^2}{4}(U_1(0)+b\partial\Psi)^2(2+\gamma_0^2), \quad d_{5/2} = U_1(0)U_1(x)\frac{\gamma_0^2}{4}(U_1(0)+b\partial\Psi)^2\gamma_0,$$

$$d_3 = U_1(0)U_1(x)\frac{\gamma_0^2}{4}(U_1(0)+b\partial\Psi)^4\frac{1}{4}.$$

Let us present one more illustration to the above considerations, which, unlike the examples in (18), (22), (33), is based on experimental results. This example clearly shows the importance of taking into account the change in entropy and the effect of external influences on the average *FPT*.

In [84, 85] the authors studied the fluctuations of the time elapsed until the electric charge transferred through a conductor reached a given threshold value. The distribution of the first-passage times for the net number of electrons transferred between two metallic islands in Coulomb blockade regime is considered. In [84] a simple analytical approximation was derived for the first-passage-time (*FPT*) distribution, which takes into account the non-Gaussian statistics of the electron transport, and showed that it is capable of describing the experimental distributions with high accuracy.

The average *FPT* is expressed in terms of the entropy change accompanying this process. In addition to the changes that are necessarily adherent to the course of a random process, any other changes in entropy can be taken into account, which correspond to some other processes in the system. The net effect of such changes affecting the average *FPT* is illustrated by taking into account different values of the bias voltage. The applied voltage acts as an outer impact on the system. The average *FPT* with a zero value of the argument of the Laplace transform of the *FPT* distribution density, which is associated with changes in the entropy of the system, does not reflect the impact of real processes on the average *FPT*. It is necessary to take into account those changes in entropy that accompany the random process of reaching a certain boundary.

Let us briefly discuss the connection between distribution (8) and superstatistics [8, 9]. There, the assumption is not made about the *FPT* distribution not depending on the random internal energy $u$; such a dependence is taken into account. A stochastic storage model [86] is used as a stochastic process for this system. The distributions of superstatistics from [8-9] can be obtained in another way, different from the approach of [8-9], if we assume that the intensity of energy flow in the system $\lambda$ depends on the density of the total random internal energy of the system $u$ as $\lambda \to \lambda_0/(n_0+u)$. The average value of one stepwise input $x_0$ into the system [the input in simple stochastic storage models is a series of Poisson distributed batch inputs with some distribution $b(x)$ of the material in each batch, see [86] for details] is cast as $\bar{x} \to \bar{x}_0(n_0+u)$. For a constant release rate equal to unity, as in the corresponding storage model [103], the partition function $Q_k$ does not change; (see for details [8, 9, 103, 86, 101]) $Q_k^{-1} = P_{0k} = 1-\rho$; $\rho = \lambda\bar{x}$, $\bar{x} = \int xb(x)dx$, $b(x)$ is distribution function one stepwise input with average value $\bar{x}$. Such substitution can be substantiated by physical considerations that with an increase in the energy of the system and an increase in the number of particles in it, it is more difficult for flows of energy and the number of particles to enter the system.

In our works [8, 9] we have obtained the superstatistics in a way different from that of [100]. This approach is free from the shortcomings pointed out in [102]. The proposed approach can be considered as an extension and refinement of the theory of superstatistics. The obtained distribution contains the new parameter related to a thermodynamic state of the system, and also with distribution of a lifetime of a metastable states and interaction of this states with an environment.

The Gibbs distribution does not describe the dissipative processes that develop in the system. Superstatistics describe systems by constantly putting energy into a system with permanent dissipation. The value $\alpha=\gamma/\lambda$ is connected with dissipative processes in the system (through parameter of distribution (8)-(9) $\gamma$). It defines the correlation between Gibbs and superstatistics multipliers in the distribution from [8, 9]. Other forms of distributions of the form of superstatistics are possible, which can be derived from a distribution of the form (8) - (9) [103].

The original distribution (9) contains two distribution functions: $f$ and $R$ (13), in contrast to superstatistics, which contain only one distribution density for the reciprocal temperature. Therefore, the distribution (9) and distributions obtained from it have greater capabilities than superstatistics. When obtaining expressions from (9) [8, 9], only the gamma distribution function for $f(T_\gamma)$ (13) was used. Another type for this distribution function can be chosen as well.